\documentclass[11pt]{article}
\pdfoutput=1
\usepackage{amsmath}
\usepackage{amssymb}
\usepackage{graphicx,bbm,mathrsfs}
\usepackage{nicefrac}
\usepackage{bbm}
\usepackage{geometry}       
\usepackage{jheppub}     

\usepackage{dcolumn}   
\usepackage{bm}        
\usepackage{graphicx,mathrsfs}
\usepackage{nicefrac}
\usepackage{multirow}
\usepackage{color}

\def\ie{{\it i.e.}}

\newcommand{\be}{\begin{equation}}  
\newcommand{\ee}{\end{equation}}  
\newcommand{\bea}{\begin{eqnarray}}  
\newcommand{\eea}{\end{eqnarray}}  

\def\gev{\, {\rm GeV}}

\newcommand{\tr}{\operatorname{tr}}

\DeclareRobustCommand{\fbi}{\ensuremath{\mathrm{fb}^{-1}}}

\begin{document}

\vspace*{1.2cm}

\begin{center}

\thispagestyle{empty}
{\Large\bf 
Anomalous light-by-light scattering at the LHC: recent developments and future perspectives
}\\[10mm]

\renewcommand{\thefootnote}{\fnsymbol{footnote}}

{\large  Sylvain~Fichet$^{\,a,b}$
\footnote{sylvain.fichet@gmail.com} }\\[10mm]

\addtocounter{footnote}{-1} 

{\it $^a$ ICTP South American Institute for Fundamental Research, Instituto de Fisica Teorica,\\
Sao Paulo State University, Brazil \\
}
{\it
$^{b}$~International Institute of Physics, UFRN, 
Av. Odilon Gomes de Lima, 1722 - Natal-RN, Brazil \\
}

\vspace*{12mm}

{  \bf  Abstract }
\end{center}

%
%

The installation of forward proton detectors at the LHC will provide the possibility to perform  high-precision measurements, opening a novel window on physics beyond the Standard Model. We review recent simulations and theoretical developments about the measurement of anomalous light-by-light scattering. 
The search for this process will provide bounds on a wide range of new particles. 
Future perspectives for precision QED at the LHC are also briefly discussed.


\noindent
\clearpage

\section{Effective Lagrangian and precision physics}\label{se:intro}

Several major facts like  the gauge-hierarchy problem or the observation of dark matter  suggest that  a new physics beyond the Standard Model of particles (SM) should emerge at a mass scale close from the electroweak  scale. However, after the first LHC run, a certain amount of popular models has 
been ruled out or they are cornered in fine-tuned regions of their parameter 
space.
While the next LHC run is coming, it is more than ever important to be 
prepared to search for any kind of new physics in the most possible robust ways. 

In a scenario of new physics out of reach from direct observation at the LHC, one may expect that the first manifestations show up in precision measurements of the  SM properties. 
Assuming that the new physics scale $\Lambda$ is higher than the typical LHC energy reach $E_{\rm LHC}$, the correlation functions of the SM fields can be expanded with respect to $E_{\rm LHC}/\Lambda$. At the Lagrangian level, this generates a series of local operators of higher dimension, which describe all the manifestations of new physics observable at low-energy.
 This low-energy effective Lagrangian reads
\be
{\cal L}_{\rm eff}={\cal L}_{\rm SM}+\sum_{i,n}\frac{\alpha_i^{(n)}}{\Lambda^n}\,.
\ee
The coefficients $\alpha_i^{(n)}$ are roughly $O(1)$ if generated at tree-level or 
$O(1/16\pi^2)$ if generated at one-loop level. 

The effective Lagrangian is somehow the natural companion of precision physics. 
In all generality, the goal of SM precision physics is to get information on the coefficients $\alpha_i^{(n)}$ and the new physics scale $\Lambda$. 
For a given set of data, bounds on $\alpha_i^{(n)}$ can be obtained if one fixes $\Lambda$. 
However, it is also obviously interesting to draw bounds on $\Lambda$ itself. 
 In order to get meaningful bounds on $\Lambda$, a  statistical subtlety has to be taken into account (see \cite{Fichet:2013jla}), that conceptually boils down to require new physics to be testable.

Among the various sectors of the SM that can be probed at the LHC, the pure Yang-Mills sector describes triple and quartic gauge boson interactions, that are all fixed by gauge symmetry. 
Among these interactions, the self-interactions of neutral gauge bosons are particularly appealing. Indeed, these interactions are generated only at loop-level in the SM, such that the SM irreducible background is small. Neutral gauge-bosons self-interactions  should be thus considered as  smoking-gun observables for new physics. 

For $\Lambda>E_{\rm LHC}$, neutral gauge-boson interactions beyond the SM are described  by dimension-8 operators with two kinds of structure, $(DH)(DH)^\dagger VV\,/\Lambda^4$ and $VVVV\,/\Lambda^4$. Schematically, the former   is expected to dominate for energies lower than the electroweak scale, while the later is expected to dominate for energies higher than the EW scale. The second kind, \ie~\textit{pure-gauge} operators, are thus fully relevant for the LHC, and should be the dominant ones at a future collider with higher energy reach.

Four-photon interactions are described by two pure-gauge operators, 
\be
\mathcal{L}_{4\gamma}= 
\zeta_1 F_{\mu\nu}F^{\mu\nu}F_{\rho\sigma}F^{\rho\sigma}
+\zeta_2 F_{\mu\nu}F^{\nu\rho}F_{\rho\lambda}F^{\lambda\mu}
\label{zetas} \,.
\ee
The effect of any   object beyond the SM can be parametrized in terms of the $\zeta_1$, $\zeta_2$ parameters, as well as any experimental search results.

\section{Precision physics with intact protons}\label{se:AFP}

\begin{figure}
\begin{center}
\includegraphics[width=0.70\linewidth]{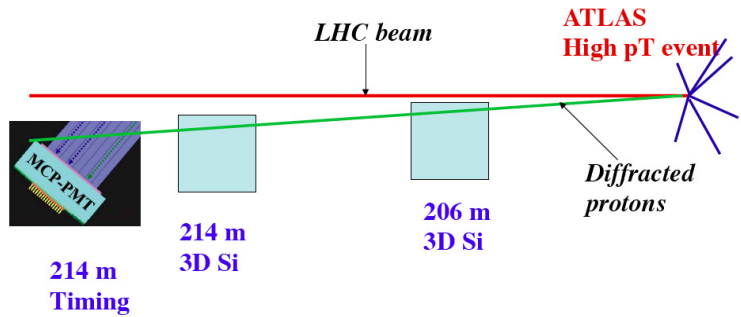}
\end{center}
\caption{Scheme of the AFP detector. Roman pot hosting Si and timing detectors
will be installed on both sides of ATLAS at 206 and 214 m from the ATLAS nominal interaction point. The
CMS-TOTEM collaborations will have similar detectors.}
\label{afp}
\end{figure}

New possibilities for precision measurements will be opened with the installation of the new forward detectors, which is scheduled at both ATLAS (ATLAS Forward Proton detector \cite{atlas}) and CMS (CT-PPS detector \cite{cms}).
The purpose of these detectors is to measure intact protons arising from diffractive processes at small angle.  They will be built at $\sim200$ m on both sides of CMS and ATLAS. The detectors should host  tracking stations, as well as timing detectors (see Fig.~\ref{afp}).
The proton taggers 
are expected to determine the fractional proton momentum loss $\xi$ in the 
range $0.015 < \xi < 0.15$ with a relative resolution of 2\%. 
In addition, 
the time-of-flight of the protons can be measured within 10~ps, which 
translates into $\sim 2$~mm resolution on the determination of the interaction 
point along the beam axis $z$. 

The crucial feature of the forward detectors is that they provide  the complete kinematics of the event, which in turn can be used to drastically reduce the backgrounds. This setup constitutes an excellent method to look for the effective operators describing physics beyond the SM. Proton scattering processes with intermediate photons are the mostly studied ones, because the equivalent photon approximation is well understood. In principle, at the LHC energies, intermediate $W$, $Z$ bosons could 
also happen, however a precise estimation of the fluxes is needed. 

\begin{figure}
\begin{center}
\includegraphics[trim=0cm 0cm 0cm 0cm, clip=true,width=9cm]{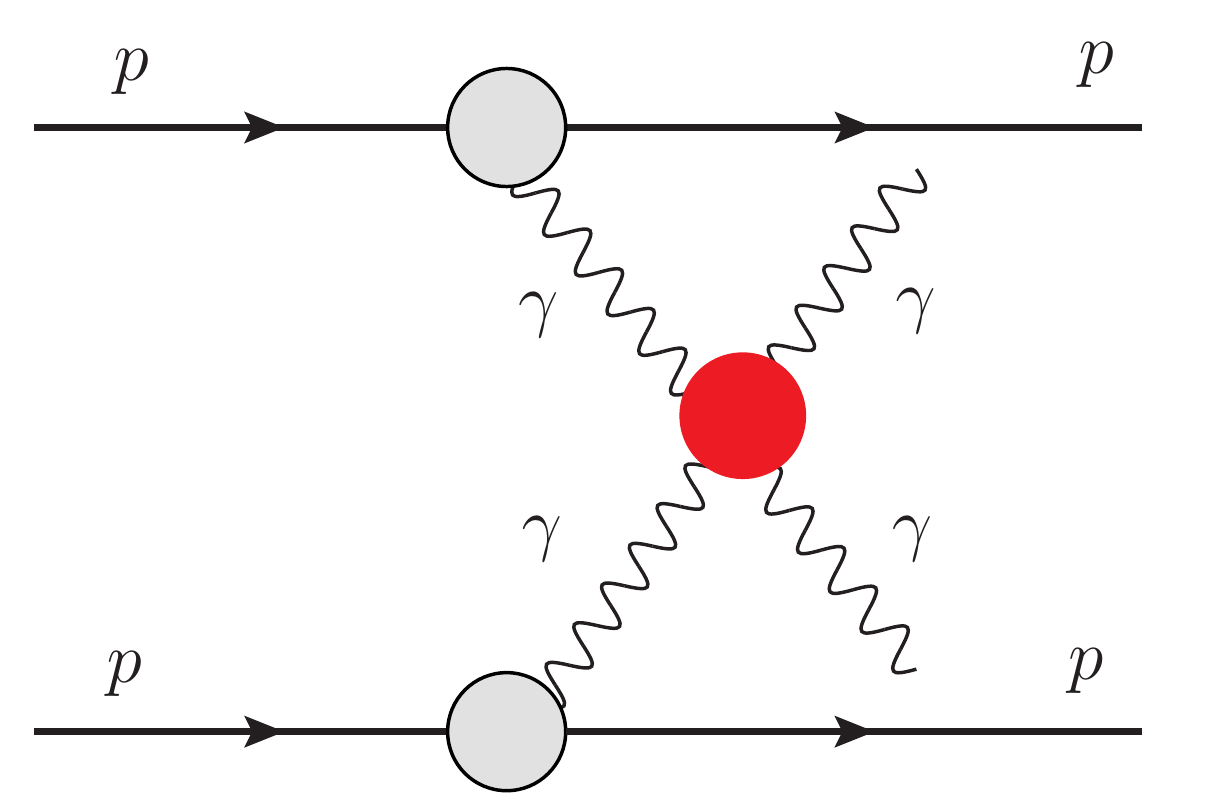}
\end{center}
\caption{Light-by-light scattering with intact protons. }
\label{fig:4gamma}
\end{figure}

Forward proton detectors open thus a new window on physics beyond the SM. They provide a clean environment to search for the effective operators describing physics beyond the SM. For example, operators like $|H|^2 V_{\mu\nu}V^{\mu\nu}/\Lambda^2$ induce anomalous  single or double Higgs production (for the MSSM case, see \cite{Heinemeyer:2007tu, Tasevsky:2014cpa}). 
The flavour-changing dipole operators like $\bar{q}\sigma_{\mu\nu} t V^{\mu\nu} /\Lambda^2$ induce single top plus one jet production (see \cite{Fichet:2015oha}). Finally, the four-photon operators of Eq.~\ref{zetas} induce light-by-light scattering. 
This last process is pictured in Fig.~\ref{fig:4gamma}.
Studies using proton-tagging at the LHC 
for new physics searches can be found in \cite{usww, usw,Sahin:2009gq,Atag:2010bh, Gupta:2011be, Lebiedowicz:2013fta, Fichet:2013ola,Fichet:2013gsa,Sun:2014qoa,
Sun:2014qba,Sun:2014ppa,Sahin:2014dua,Inan:2014mua,Fichet:2014uka}.

\label{se:fwd}

\section{Light-by-light scattering at the LHC}\label{se:simu}

Given the promising possibilities of forward detectors, a  realistic simulation of the search for anomalous $\gamma\gamma \rightarrow  \gamma\gamma$ at the $14$ TeV LHC has been carried out in \cite{Fichet:2014uka}. 
The search for light-by-light scattering at the LHC without proton tagging  has been first thoroughly analyzed in \cite{friend:2013yra}.
Let us review the setup, the backgrounds, the event selection, and the sensitivity to the $\zeta_{1,2}$ anomalous couplings expected at the $14$ TeV LHC.

The Forward Physics Monte Carlo  generator (FPMC, \cite{FPMC})  is designed to produce within a same framework the double pomeron exchange (DPE), single diffractive,
exclusive diffractive and photon-induced processes. 
The emission of photons by protons is correctly described by 
the  Budnev flux \cite{Chen:1973mv, Budnev:1974de}, which takes into account the proton electromagnetic structure.
The SM $\gamma\gamma \rightarrow  \gamma\gamma$  process induced by loops of SM fermions and $W$, the exact contributions from new particles with  arbitrary charge and mass, and 
 the anomalous vertices described  by the effective operators Eq.~\eqref{zetas} have been implemented into FPMC.

The backgrounds are divided into three classes. Exclusive 
processes with two intact photons and a pair of photon candidates include the 
SM light-by-light scattering, $\gamma\gamma \rightarrow e^+e^-$ and the central-exclusive production of two 
photons via two-gluon exchange, simulated using ExHume~\cite{ExHuME}.
Processes involving DPE can result in protons 
accompanied by two jets, two photons and a Higgs boson that decay into two 
photons. Finally, one can have gluon or quark-initiated production of two 
photons, two jets or two electrons (Drell-Yan) with intact protons arising 
from  pile-up interactions.

The knowledge of the full event kinematics is a powerful constraint
to reject the  background from pile-up. The crucial cuts consist in matching the missing momentum (rapidity difference) of the di-proton system  with the invariant mass (rapidity difference) of the di-photon system, which is measured in the central detector.  Extra cuts rely on the event topology, using the fact that the photons are emitted back-to-back with similar $p_T$. Further background reduction could even possible by  measuring the protons
time-of-flight, which provides a complete reconstruction of the primary vertex with a typical precision of 1mm.


The estimation of  the LHC sensitivities to effective four-photon couplings $\zeta_i$ provided by measuring  light-by-light scattering with proton tagging is performed in \cite{Fichet:2013gsa,Fichet:2014uka}. 
These sensitivities  are given in Table~\ref{sensitivities}
for different scenarios corresponding to the medium luminosity at the LHC
(300 fb$^{-1}$) and the high luminosity (3000 fb$^{-1}$ in ATLAS). The $5 \sigma$ discovery potential as well as the
95\% CL limits with a pile-up of 50 are given.

It turns out that the selection efficiency is sufficiently good so that the background amplitudes are negligible with respect to the anomalous $\gamma\gamma\rightarrow\gamma\gamma$ signal. A handful of events is therefore enough to reach a high significance. 
In that regime, the signal-background interference can be neglected, and the unpolarized differential cross-section in presence of effective operators takes a simple form
\be
  \frac{d\sigma}{d\Omega}
  =\frac{1}{16 \pi^2\,s}(s^2+t^2+st)^2
  \left[48 (\zeta_1)^2 + 40 \zeta_1 \zeta_2 + 11 (\zeta_2)^2\right]
  \label{xsec}
  \ee
where $s$, $t$ are the usual Mandelstam variables. 

 The obvious inconvenience of the EFT approach is that it is valid only in the high mass region, $m\gg E$. In order to use the EFT result down to $m\sim E$, it is common to introduce ad-hoc form factors which mimics the behaviour of the -- unknown -- amplitudes near the threshold. 
Clearly, this method introduces  arbitrariness into the results. Not only do the results depend on the functional form of the form factor, but also on the energy scale at which they are introduced.

\begin{table}

\begin{center}
\begin{tabular}{|c||c|c||c|}
\hline
Luminosity &  300~\fbi & 300~\fbi & 3000~\fbi \\
\hline
 pile-up ($\mu$)  & 50 & 50 & 200 \\
\hline
\hline
coupling (GeV$^{-4}$) &  5 $\sigma$ & 95\% CL & 95\% CL \\
\hline
$\zeta_1$&   $1.5\cdot10^{-14}$ & $9\cdot10^{-15}$  & $7\cdot10^{-15}$\\
\hline
$\zeta_2$&   $3\cdot10^{-14}$& $2\cdot10^{-14}$  & $1.5\cdot10^{-14}$ \\
\hline

\end{tabular}
\end{center}

\caption{5\,$\sigma$ discovery and 95\% CL exclusion limits on $\zeta_1$ and $\zeta_2$ 
couplings in\gev$^{-4}$ (see Eq.~\ref{zetas}). All sensitivities are given for 300 fb$^{-1}$
and $\mu=50$  pile-up events (medium luminosity LHC) except for the numbers of the last column which are given for 3000
fb$^{-1}$ and $\mu=200$  pile-up events (high luminosity LHC). }
\label{sensitivities}

\end{table}

\section{ Sensitivity to generic charged particles }
\label{se:lhc}

What about actual new physics candidates ?
The perturbative contributions to anomalous gauge couplings appear at one-loop and 
 can be parametrized in terms of the mass and quantum numbers of the new particle \cite{Fichet:2013ola}. In the case of four-photon interactions, only electric charge matters. 
 New  particles with exotic electric charges can for example appear 
in composite Higgs model \cite{Agashe:2004rs} or in warped extra-dimension models with custodial symmetry \cite{Agashe:2003zs}.
 The new particles  have  in general a 
multiplicity with respect to electromagnetism. For instance, the multiplicity is
three if the particles are colored. 
It is convenient to take into account this multiplicity by defining
\be
Q_{\rm eff}^4=\tr Q^4
\ee
where the trace goes over all particles with the same approximate mass. 
  In the case of new electrically charged particles with arbitrary spin $S$, 
 the coefficients read
\be 
\zeta_i=\frac{\alpha^2_{\rm em} Q_{\rm eff}^4}{m^{4}}\,c_{i,S}\,,
\label{zetaEFT}
\ee 
where
\be
c_{1,S}=
\begin{cases}
\frac{1}{288} & S=0 \\
-\frac{1}{36} & S=\frac{1}{2} \\
-\frac{5}{32} & S=1 \\
\end{cases}
\,,\quad
c_{2,S}=
\begin{cases}
\frac{1}{360} & S=0 \\
\frac{7}{90} & S=\frac{1}{2} \\
\frac{27}{40} & S=1  \\
\end{cases} \quad.
\label{EH}
\ee
  The contributions from the scalar are smaller by one order of magnitude  
  with respect to the fermion and vector.
It can easily be checked that in the case of fermions $\mathcal L_{4\gamma}$ reduces to the famous Euler-Heisenberg Lagrangian\cite{Heisenberg:1935qt}. \footnote{These results also match early computations \cite{Boudjema:1986rh,Baillargeon:1995dg}}.

The effective field theory analysis has the advantage of being very simple. 
However it is only  
 valid as long as the center-of-mass energy is small with respect to the 
 threshold of pair-production of real particles, $s\ll 4m^2$. 
Since the maximum proton missing mass (corresponding to the di-photon invariant mass in our case) is of the order of $\sim 2$ TeV at the 14 TeV LHC, for particles lighter than $\sim 1$ TeV the  effective field theory computation needs to be corrected.
This can be done by using ad-hoc form factors, as often done in the literature. 
The more correct approach is to take into account the full momentum dependence of the four-photon amplitudes. The SM loops have been computed in Refs.~\cite{Karplus:1950zza,Karplus:1950zz,Costantini:1971cj,Jikia:1993tc} and are collected in Ref. \cite{Fichet:2014uka}. At LHC energies, the $W$ loop dominates over all fermion loops including the top because it grows logarithmically.

\begin{figure}
\begin{center}
\includegraphics[width=0.49\linewidth]{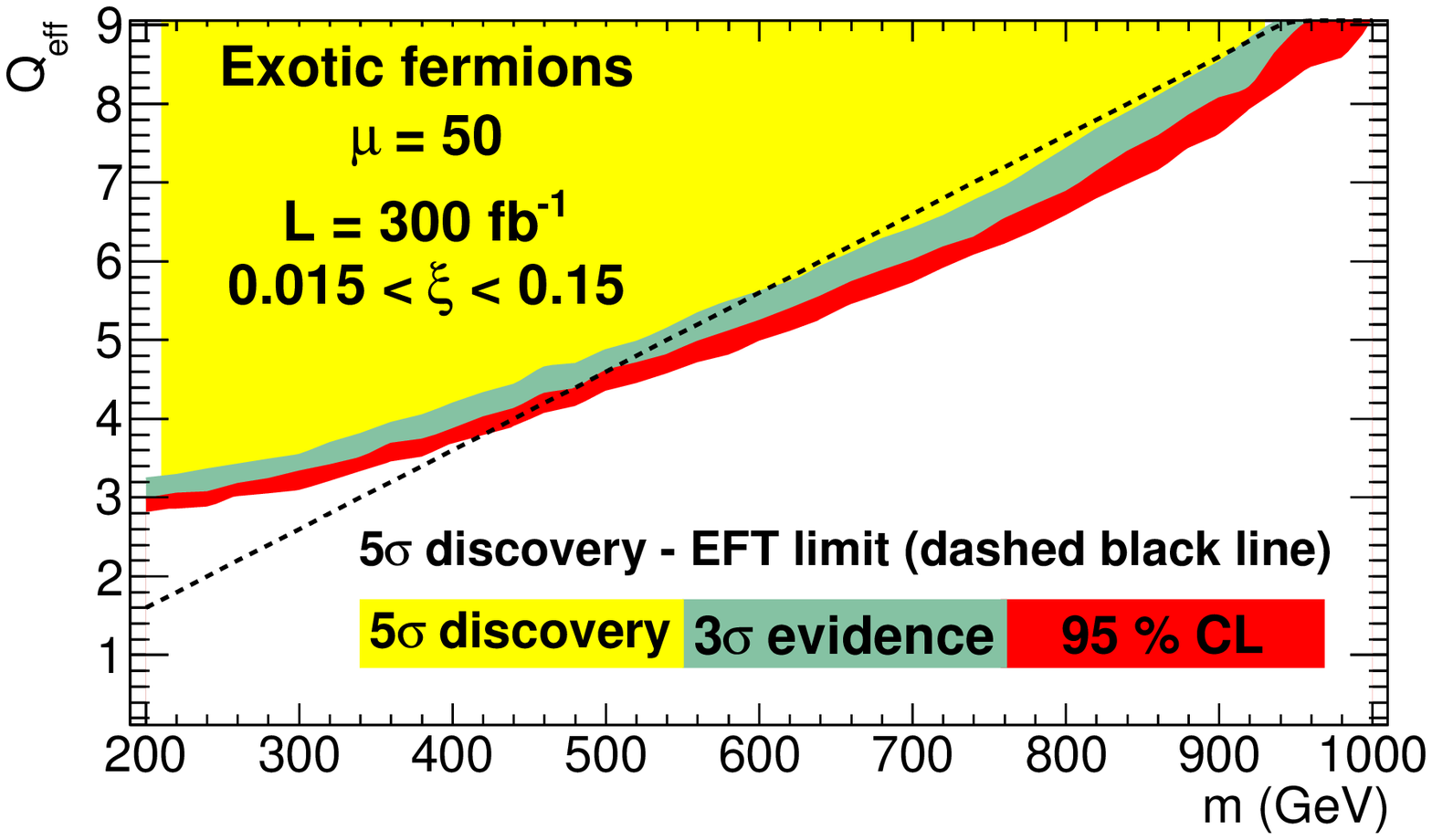}
\includegraphics[width=0.49\linewidth]{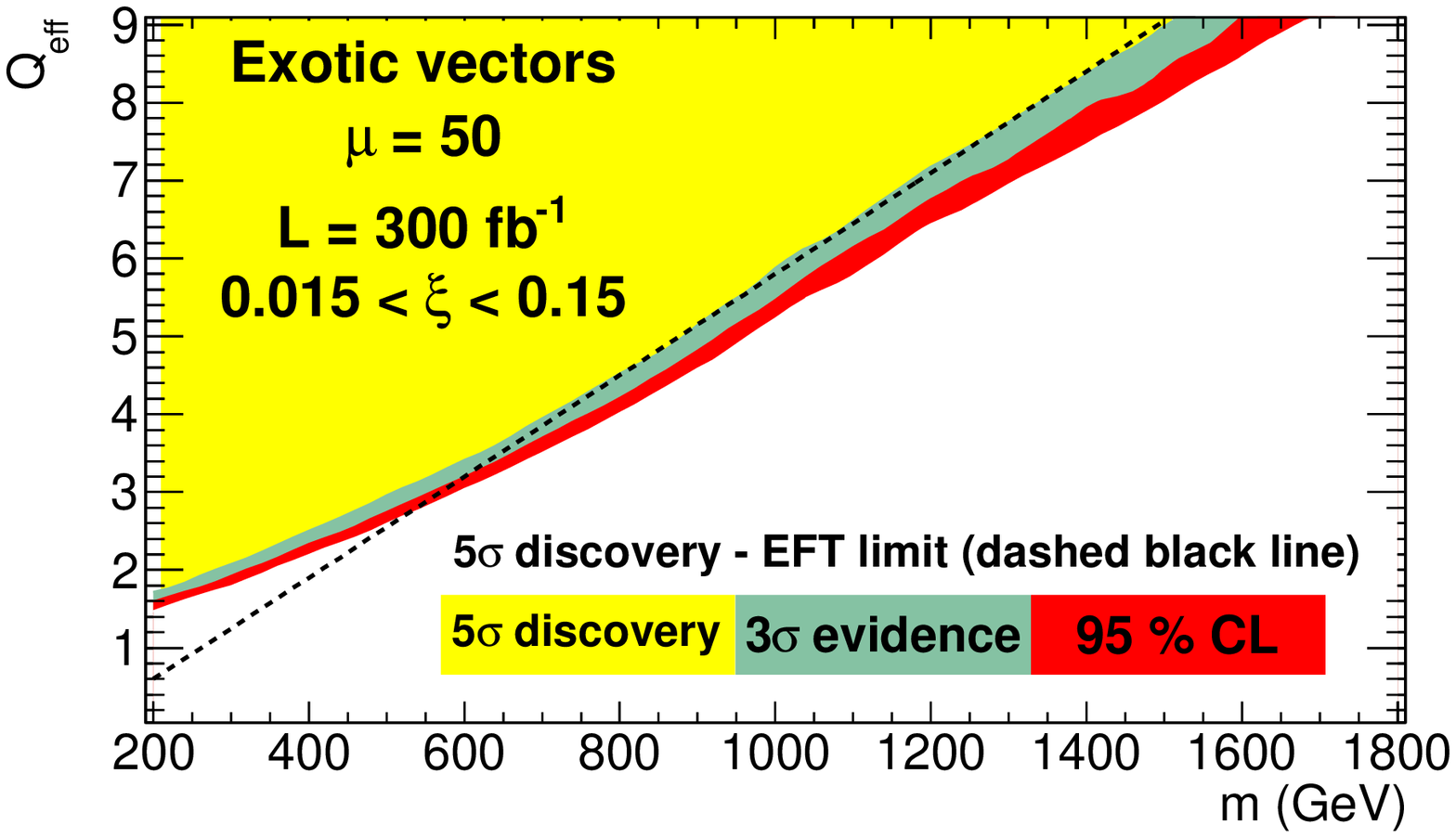}
\end{center}
\caption{Exclusion plane in terms of mass and effective charge of generic fermions and vectors with full integrated luminosity at the medium-luminosity LHC (300~\fbi, $\mu=50$).}
\label{fig:mqplane}
\end{figure}

The  results of the simulation with full amplitudes are given in Tab. \ref{fullamp_values} and
Fig.~\ref{fig:mqplane}  where are displayed the 5$\sigma$ discovery, 3$\sigma$
evidence and 95\% C.L.~limit for fermions and vectors 
for a luminosity of 300 fb$^{-1}$ and a pile-up of 50. 
It is found  that a vector (fermion) with  $Q_{\rm eff}=4$, can be discovered up to mass $m=700$~GeV ($370$~GeV). At high mass, the exclusion bounds follow isolines $Q\propto m$, as dictated by the EFT couplings Eq. \ref{zetaEFT}. Extrapolating the same analysis to a higher luminosity of 3000 \fbi for a pile-up of 200 leads to a slighlty improved sensitivity of  $m = 740$~GeV ($410$~GeV) for vectors (fermions).

One may notice that some searches for vector-like quarks, as motivated from e.g.~Composite Higgs models, already lead to stronger bounds than the ones projected here. For instance, vector-like top partners arising from the $(2,2)$ (corresponding to $Q_{\rm eff}\approx2.2$) of mass $m=500$ GeV would be excluded from present LHC data, while they would be out of reach using light-by-light scattering.
On the other hand, the light-by-light scattering results are completely model-independent. They apply 
just as well to different effective charges, are independent of the amount 
of mixing with the SM quarks, and even apply to vector-like leptons!

Four-photon amplitudes also contribute to  the magnetic dipole moment of the muon $a_\mu$ via two and three-loop diagrams. An estimating of these loop contributions shows that with an experimental bound on
 $a_\mu\sim 6\cdot 10^{-10}$, the sensitivity of this measurement is $m/Q_{\rm eff}\sim 5$ GeV. 
Comparing this estimate to the projections from Fig.~\ref{fig:mqplane}, it appears that, despite its impressive accuracy, the $g-2$ measurement is not 
competitive with the light-by-light scattering measurement.

\begin{table}

\begin{center}
\begin{tabular}{|c||c|c|c|c|c|}
\hline
Mass (GeV) & 300 & 600 & 900 & 1200 & 1500 \\
\hline
$Q_{\rm eff}$ (vector)  & 2.2 & 3.4 & 4.9 & 7.2 & 8.9 \\
\hline
$Q_{\rm eff}$ (fermion) & 3.6 & 5.7 & 8.6 & - & - \\
\hline
\end{tabular}
\end{center}

\caption{5\,$\sigma$ discovery limits on the effective charge of new generic charged fermions and vectors for various masses scenarios and full integrated luminosity at the medium-luminosity LHC (300~\fbi, $\mu=50$).}
\label{fullamp_values}

\end{table}

\section{Sensitivity to neutral particles}\label{se:charged}

\begin{figure}
\begin{picture}(400,250)
\put(50,0){		\includegraphics[trim=0cm 0cm 0cm 0cm, clip=true,width=10cm]{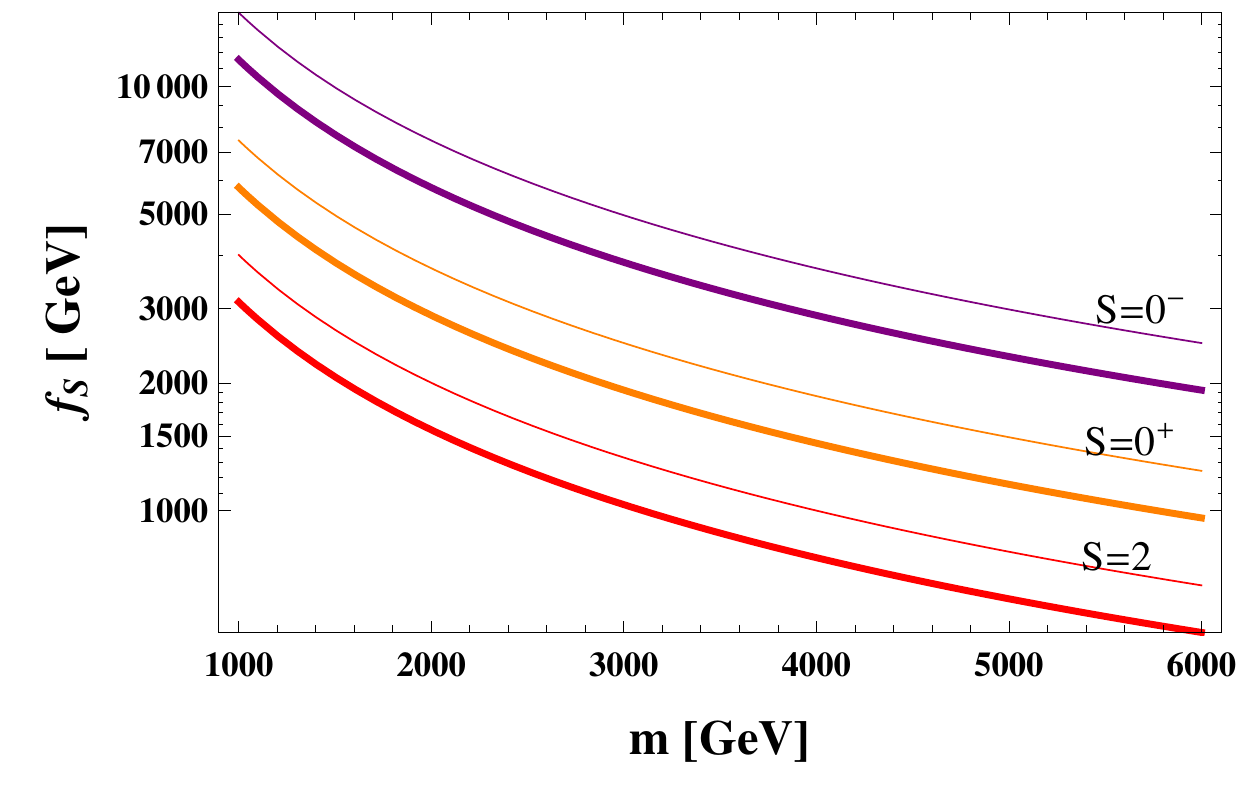}}
\end{picture}
\vspace{0.5cm}
	\caption{ Sensitivities for the neutral simplified models in the $(m,f_S)$ plane. 	
	Thick lines correspond to 5\,$\sigma$, thin lines correspond to 95\% CL limits. 
The limits are given for the medium luminosity LHC with all photons (no conversion required) and no form-factor (see Tab. \ref{sensitivities}).		
	 \label{fig:fm_plot}}
	\end{figure}

Beyond the perturbative contributions from  
charged particles,  non-renormalizable interactions of neutral particles are 
also present in common extensions of the SM.  Such theories can contain 
scalar, pseudo-scalar and spin-2 resonances, respectively denoted by $\varphi$, 
$\tilde \varphi$ and $h^{\mu\nu}$~\cite{Fichet:2013gsa}. Independently of the particular new physics model they originate from, their leading couplings to the photon  
are fixed completely by Lorentz and CP symmetry as 
\be\begin{split}
\mathcal L_{\gamma\gamma}=&f_{0^+}^{-1}\,\varphi\, 
(F_{\mu\nu})^2+f_{0^-}^{-1}\, \tilde\varphi \, F_{\mu\nu}F_{\rho\lambda}\,
\epsilon^{\mu\nu\rho\lambda} \\&+f_{2}^{-1}\, h^{\mu\nu}\, (-F_{\mu\rho} 
F_{\nu}^{\,\,\rho}+\eta_{\mu\nu} (F_{\rho\lambda})^2/4)\,,
\end{split}
\ee
where the $f_S$ have mass dimension 2. 
They then generate  $4\gamma$ couplings by tree-level exchange as 
$\zeta_i=(f_{S}\, m)^{-2}\,d_{i, s}$, where
\be
d_{1,s}=
\begin{cases}
\frac{1}{2} & s=0^+ \\
 -4 & s=0^- \\
-\frac{1}{8} & s=2\\
\end{cases}
\,,\quad
d_{2,s}=
\begin{cases}
0 & s=0^+ \\
8  & s=0^- \\
\frac{1}{2} & s=2 \\
\end{cases}
\,.
\ee
The model independent sensitivities for these three cases are shown in Fig.~\ref{fig:fm_plot}.

It appears that the non-renormalizable contributions from neutral particles are sensibly larger than the charged particles contributions. Light-by-light scattering offers therefore  a privileged window on strongly interacting phenomena. Considering actual models, two kind of candidates are known: the Kaluza-Klein (KK) gravitons and the strongly-interacting heavy dilaton (SIHD). 
\begin{itemize}
\item {\it Kaluza-Klein gravitons:}
 The contribution of the entire tower of KK gravitons of warped extra dimensions   is computed in \cite{Fichet:2013ola}. The strength of warped gravity $\kappa$ can be taken of order unity.
For $\kappa=2$, and using the 5\,$\sigma$ and 95\% CL sensitivities   for the medium luminosity LHC  (see Tab. \ref{sensitivities}), the effect of the KK resonances can be detected up to  mass 
\be
m_{\rm KK}<5670\, \textrm{GeV} \,(5\,\sigma)\,,\quad m_{\rm KK}<6450\, \textrm{GeV}\,\,(95\%{\rm CL})\,.
\ee	
These sensitivities are competitive with respect to searches for direct production of KK resonances at the LHC.

\item{\it Strongly-interacting dilaton}~\cite{Fichet:2014uka}:
Extensions of the Standard Model sometimes feature a new strongly-interacting sector. Provided that this sector is conformal in the UV, it is most likely explicitly broken in the IR, at least by the appearance of electroweak scale and QCD confinement. As a result, the spectrum of the strong sector features a neutral scalar, the so-called dilaton, whose mass lies close to the scale of conformal breaking. In the absence of fine-tuning the dilaton's couplings are unsuppressed with respect to this scale. To distinguish it from the weakly coupled (fine-tuned) light dilaton often considered in the literature one refers to it as the Strongly-Interacting Heavy Dilaton. If the photon is at least partially composite, it also couples strongly to the dilaton. 
Using the 5\,$\sigma$ and 95\% CL sensitivities for the medium luminosity LHC, the effect of the SIHD can be detected up to mass
\be
m_{\varphi}<4260\, \textrm{GeV}\,(5\,\sigma)\,,\quad m_{\varphi}<4840\, \textrm{GeV}\, (95\%{\rm CL})\,.
\ee	
\end{itemize}

\begin{figure}[t]
\begin{picture}(400,250)
\put(50,0){		\includegraphics[trim=0cm 0cm 0cm 0cm, clip=true,width=10cm]{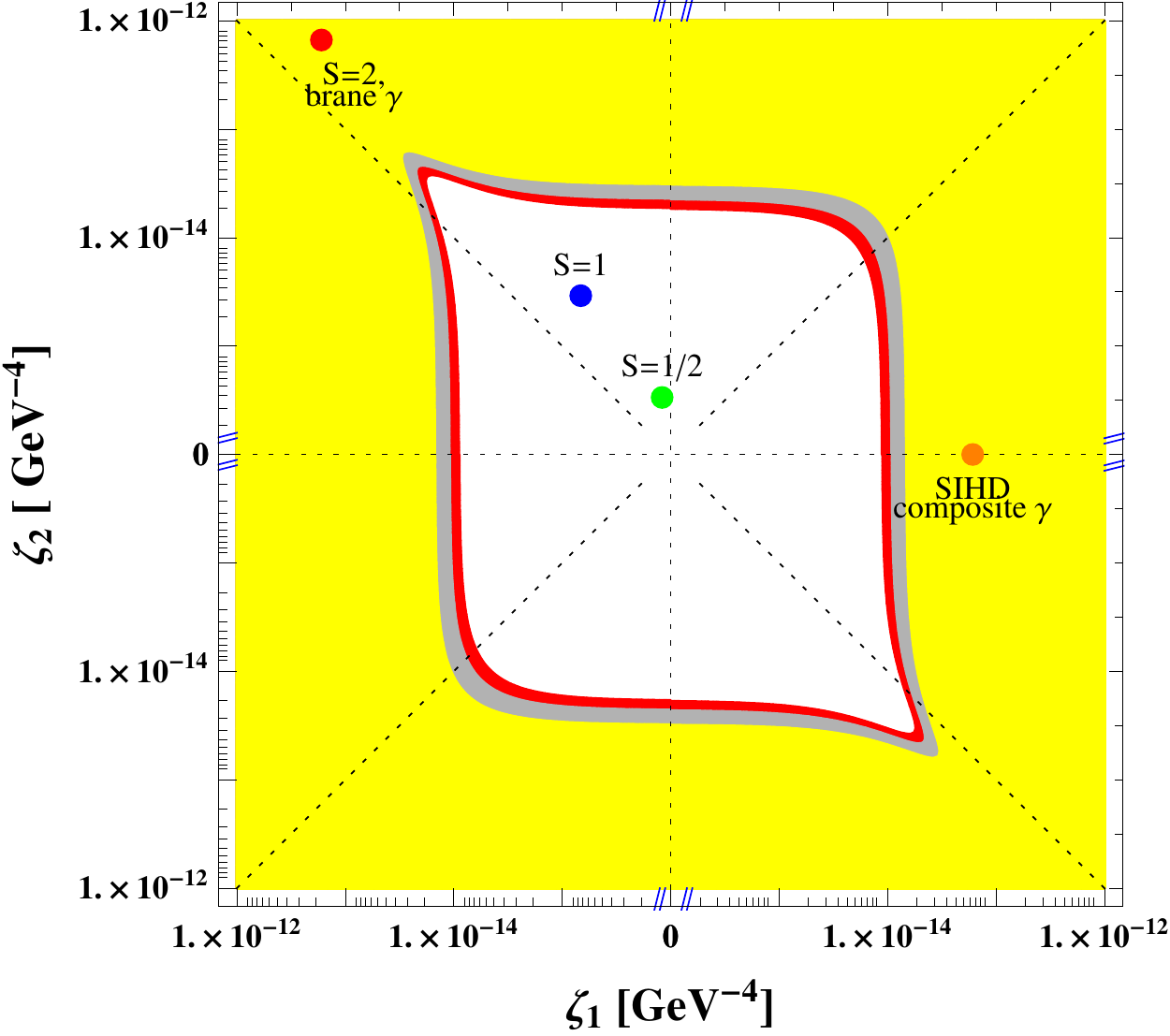}}
\end{picture}
\vspace{0.5cm}
	\caption{ Experimental sensitivity and models in the $(\zeta_1,\zeta_2)$ plane. Axes follow a logarithmic scale spanning $|\zeta_i|\in[10^{-12},10^{-16}]$.
	The yellow, grey, and red  regions can be probed at 5\,$\sigma$, 3\,$\sigma$ and 95\%~CL using proton tagging at the LHC, while 
	the white region remains inaccessible. 	The limits are given for the medium luminosity LHC with all photons (no conversion required) and no form-factor (see Tab. \ref{sensitivities}).	
	Also shown are contributions from electric particles with spin $1/2$ 
	and $1$, charge $Q_{\rm eff}=3$, mass $m=1$ TeV,  
	the contribution from warped KK gravitons with mass $m_{\rm KK}=3$ TeV, 
	$\kappa=2$ and brane-localized photon, and the
	contribution from a strongly-interacting heavy dilaton (SIHD) 
	with mass $m_{\varphi}=3$ TeV coupled to a composite photon.
	 \label{fig:zeta_plot}}
	\end{figure}

\section{Summary and perspectives}

The installation of forward proton detectors at the LHC will provide a -- somewhat surprising -- opportunity to measure the scattering of light by light, providing a new window on physics beyond the Standard Model. Recent simulations and theoretical developments show that such precision measurement  gives access to a wide range of new particles, both electrically charged and neutral. A summary plot with the expected sensitivity at the $14$ TeV LHC as well as new physics candidates is shown in Fig. \ref{fig:zeta_plot}.




These positive results on precision QED at the LHC  open new perspectives, as well as new challenges, from both theoretical and experimental sides. Here is a non-exhaustive list of  works in progress and future directions.
\begin{itemize}

\item Anomalous three-photon production. The $4\gamma$ operators contribute to anomalous $\bar q q\rightarrow \gamma\gamma\gamma$ production. Contrary to the light-by-light scattering case, one photon is virtual. 
It is interesting to evaluate the sensitivity provided by this potential measurement.

\item Light-by-light scattering in heavy-ions collisions. The photon fluxes from heavy ions are coherent, and therefore enhanced by $Z^2$. On the other hand the typical center-of-mass energy of the diphoton system is smaller. It is interesting to evaluate the sensitivity provided by this potential measurement. For an earlier study, see \cite{friend:2013yra}.

\item Experimentally disentangling between $\zeta_1$ and $\zeta_2$. Polarization-based observables could play this role. This would open the possibility of identifying the nature of the new particle producing light-by-light scattering.

\item Modelling the $W$, $Z$ fluxes. At high energy, gauge boson fluxes from electroweak charges inside the nucleons can be expected to be partly coherent. Having a model of these fluxes would certainly be useful to study electroweak ultrapheripheral collisions.

\item Light-by-light scattering from higher-spin particles. 
Extended objects of higher spin do exist in many extensions of the SM. This is potentially the case with the composite states  from any strongly coupled sector, and also with the excitations of low-energy strings. The tools necessary to handle quantum computations involving higher-spin particles are under development.

\end{itemize}

\bibliographystyle{JHEP} 

\bibliography{qgc_JHEP}

\providecommand{\href}[2]{#2}\begingroup\raggedright\begin{thebibliography}{10}

\bibitem{Fichet:2013jla}
S.~Fichet, {\it {Probing the scale of New Physics at the LHC: The example of
  Higgs data}},  {\em Nucl.Phys.} {\bf B884} (2014) 379--395,
  [\href{http://arxiv.org/abs/1307.0544}{{\tt arXiv:1307.0544}}].

\bibitem{atlas}
{ATLAS collaboration, CERN-LHCC-2011-012}, {\it {Letter of intent, Phase-I
  upgrade}}, .

\bibitem{cms}
{CMS and TOTEM collaboration, CERN-LHCC-2014-021}, {\it {CMS-TOTEM Precision
  Proton Spectrometer}}, .

\bibitem{Heinemeyer:2007tu}
S.~Heinemeyer, V.~Khoze, M.~Ryskin, W.~Stirling, M.~Tasevsky, et~al., {\it
  {Studying the MSSM Higgs sector by forward proton tagging at the LHC}},  {\em
  Eur.Phys.J.} {\bf C53} (2008) 231--256,
  [\href{http://arxiv.org/abs/0708.3052}{{\tt arXiv:0708.3052}}].

\bibitem{Tasevsky:2014cpa}
M.~Tasevsky, {\it {Review of Central Exclusive Production of the Higgs Boson
  Beyond the Standard Model}},  {\em Int.J.Mod.Phys.} {\bf A29} (2014), no.~28
  1446012, [\href{http://arxiv.org/abs/1407.8332}{{\tt arXiv:1407.8332}}].

\bibitem{Fichet:2015oha}
S.~Fichet, B.~Herrmann, and Y.~Stoll, {\it {Tasting the SU(5) nature of
  supersymmetry at the LHC}},  \href{http://arxiv.org/abs/1501.0530}{{\tt
  arXiv:1501.0530}}.

\bibitem{usww}
E.~Chapon, C.~Royon, and O.~Kepka, {\it {Anomalous quartic W W gamma gamma, Z Z
  gamma gamma, and trilinear WW gamma couplings in two-photon processes at high
  luminosity at the LHC}},  {\em Phys.Rev.} {\bf D81} (2010) 074003,
  [\href{http://arxiv.org/abs/0912.5161}{{\tt arXiv:0912.5161}}].

\bibitem{usw}
O.~Kepka and C.~Royon, {\it {Anomalous $W W \gamma$ coupling in photon-induced
  processes using forward detectors at the LHC}},  {\em Phys.Rev.} {\bf D78}
  (2008) 073005, [\href{http://arxiv.org/abs/0808.0322}{{\tt
  arXiv:0808.0322}}].

\bibitem{Sahin:2009gq}
I.~Sahin and S.~Inan, {\it {Probe of unparticles at the LHC in exclusive two
  lepton and two photon production via photon-photon fusion}},  {\em JHEP} {\bf
  0909} (2009) 069, [\href{http://arxiv.org/abs/0907.3290}{{\tt
  arXiv:0907.3290}}].

\bibitem{Atag:2010bh}
S.~Atag, S.~Inan, and I.~Sahin, {\it {Extra dimensions in ${\gamma\gamma
  \rightarrow \gamma\gamma}$ process at the CERN-LHC}},  {\em JHEP} {\bf 1009}
  (2010) 042, [\href{http://arxiv.org/abs/1005.4792}{{\tt arXiv:1005.4792}}].

\bibitem{Gupta:2011be}
R.~S. Gupta, {\it {Probing Quartic Neutral Gauge Boson Couplings using
  diffractive photon fusion at the LHC}},  {\em Phys.Rev.} {\bf D85} (2012)
  014006, [\href{http://arxiv.org/abs/1111.3354}{{\tt arXiv:1111.3354}}].

\bibitem{Lebiedowicz:2013fta}
P.~Lebiedowicz, R.~Pasechnik, and A.~Szczurek, {\it {Search for technipions in
  exclusive production of diphotons with large invariant masses at the LHC}},
  {\em Nucl.Phys.} {\bf B881} (2014) 288--308,
  [\href{http://arxiv.org/abs/1309.7300}{{\tt arXiv:1309.7300}}].

\bibitem{Fichet:2013ola}
S.~Fichet and G.~von Gersdorff, {\it {Anomalous gauge couplings from composite
  Higgs and warped extra dimensions}},  {\em JHEP03(2014)102} (2013)
  [\href{http://arxiv.org/abs/1311.6815}{{\tt arXiv:1311.6815}}].

\bibitem{Fichet:2013gsa}
S.~Fichet, G.~von Gersdorff, O.~Kepka, B.~Lenzi, C.~Royon, et~al., {\it
  {Probing new physics in diphoton production with proton tagging at the Large
  Hadron Collider}},  \href{http://arxiv.org/abs/1312.5153}{{\tt
  arXiv:1312.5153}}.

\bibitem{Sun:2014qoa}
H.~Sun, {\it {Probe anomalous tqγ couplings through single top photoproduction
  at the LHC}},  {\em Nucl.Phys.} {\bf B886} (2014) 691--711,
  [\href{http://arxiv.org/abs/1402.1817}{{\tt arXiv:1402.1817}}].

\bibitem{Sun:2014qba}
H.~Sun, {\it {Large Extra Dimension effects through Light-by-Light Scattering
  at the CERN LHC}},  {\em Eur.Phys.J.} {\bf C74} (2014) 2977,
  [\href{http://arxiv.org/abs/1406.3897}{{\tt arXiv:1406.3897}}].

\bibitem{Sun:2014ppa}
H.~Sun, {\it {Dark Matter Searches in Jet plus Missing Energy in $\rm \gamma p$
  collision at CERN LHC}},  {\em Phys.Rev.} {\bf D90} (2014) 035018,
  [\href{http://arxiv.org/abs/1407.5356}{{\tt arXiv:1407.5356}}].

\bibitem{Sahin:2014dua}
I.~Sahin, M.~Koksal, S.~Inan, A.~Billur, B.~Sahin, et~al., {\it {Graviton
  production through photon-quark scattering at the LHC}},
  \href{http://arxiv.org/abs/1409.1796}{{\tt arXiv:1409.1796}}.

\bibitem{Inan:2014mua}
S.~Inan, {\it {Dimension-six anomalous $tq\gamma$ couplings in $\gamma \gamma$
  collision at the LHC}},  \href{http://arxiv.org/abs/1410.3609}{{\tt
  arXiv:1410.3609}}.

\bibitem{Fichet:2014uka}
S.~Fichet, G.~von Gersdorff, B.~Lenzi, C.~Royon, and M.~Saimpert, {\it
  {Light-by-light scattering with intact protons at the LHC: from Standard
  Model to New Physics}},  {\em JHEP} {\bf 1502} (2015) 165,
  [\href{http://arxiv.org/abs/1411.6629}{{\tt arXiv:1411.6629}}].

\bibitem{friend:2013yra}
D.~d'Enterria and G.~G. da~Silveira, {\it {Observing light-by-light scattering
  at the Large Hadron Collider}},  {\em Phys.Rev.Lett.} {\bf 111} (2013)
  080405, [\href{http://arxiv.org/abs/1305.7142}{{\tt arXiv:1305.7142}}].

\bibitem{FPMC}
M.~Boonekamp, A.~Dechambre, V.~Juranek, O.~Kepka, M.~Rangel, et~al., {\it
  {FPMC: A Generator for forward physics}},
  \href{http://arxiv.org/abs/1102.2531}{{\tt arXiv:1102.2531}}.

\bibitem{Chen:1973mv}
M.-S. Chen, I.~Muzinich, H.~Terazawa, and T.~Cheng, {\it {Lepton pair
  production from two-photon processes}},  {\em Phys.Rev.} {\bf D7} (1973)
  3485--3502.

\bibitem{Budnev:1974de}
V.~Budnev, I.~Ginzburg, G.~Meledin, and V.~Serbo, {\it {The Two photon particle
  production mechanism. Physical problems. Applications. Equivalent photon
  approximation}},  {\em Phys.Rept.} {\bf 15} (1975) 181--281.

\bibitem{ExHuME}
J.~Monk and A.~Pilkington, {\it {ExHuME: A Monte Carlo event generator for
  exclusive diffraction}},  {\em Comput.Phys.Commun.} {\bf 175} (2006)
  232--239, [\href{http://arxiv.org/abs/hep-ph/0502077}{{\tt hep-ph/0502077}}].

\bibitem{Agashe:2004rs}
K.~Agashe, R.~Contino, and A.~Pomarol, {\it {The Minimal composite Higgs
  model}},  {\em Nucl.Phys.} {\bf B719} (2005) 165--187,
  [\href{http://arxiv.org/abs/hep-ph/0412089}{{\tt hep-ph/0412089}}].

\bibitem{Agashe:2003zs}
K.~Agashe, A.~Delgado, M.~J. May, and R.~Sundrum, {\it {RS1, custodial isospin
  and precision tests}},  {\em JHEP} {\bf 0308} (2003) 050,
  [\href{http://arxiv.org/abs/hep-ph/0308036}{{\tt hep-ph/0308036}}].

\bibitem{Heisenberg:1935qt}
W.~Heisenberg and H.~Euler, {\it {Consequences of Dirac's theory of
  positrons}},  {\em Z.Phys.} {\bf 98} (1936) 714--732,
  [\href{http://arxiv.org/abs/physics/0605038}{{\tt physics/0605038}}].

\bibitem{Boudjema:1986rh}
F.~Boudjema, {\it {The Scattering of Light by Light in the Nonlinear Gauge}},
  {\em Phys.Lett.} {\bf B187} (1987) 362.

\bibitem{Baillargeon:1995dg}
M.~Baillargeon, F.~Boudjema, E.~Chopin, and V.~Lafage, {\it {New physics with
  three photon events at LEP}},  {\em Z.Phys.} {\bf C71} (1996) 431--442,
  [\href{http://arxiv.org/abs/hep-ph/9506396}{{\tt hep-ph/9506396}}].

\bibitem{Karplus:1950zza}
R.~Karplus and M.~Neuman, {\it {Non-Linear Interactions between Electromagnetic
  Fields}},  {\em Phys.Rev.} {\bf 80} (1950) 380--385.

\bibitem{Karplus:1950zz}
R.~Karplus and M.~Neuman, {\it {The scattering of light by light}},  {\em
  Phys.Rev.} {\bf 83} (1951) 776--784.

\bibitem{Costantini:1971cj}
V.~Costantini, B.~De~Tollis, and G.~Pistoni, {\it {Nonlinear effects in quantum
  electrodynamics}},  {\em Nuovo Cim.} {\bf A2} (1971) 733--787.

\bibitem{Jikia:1993tc}
G.~Jikia and A.~Tkabladze, {\it {Photon-photon scattering at the photon linear
  collider}},  {\em Phys.Lett.} {\bf B323} (1994) 453--458,
  [\href{http://arxiv.org/abs/hep-ph/9312228}{{\tt hep-ph/9312228}}].

\end{thebibliography}\endgroup

\end{document}